\documentclass{article}
\usepackage[utf8]{inputenc}
\usepackage{graphicx}
\usepackage{amsmath}
\usepackage{url}
\usepackage[colorlinks,bookmarksopen,bookmarksnumbered,citecolor=black,urlcolor=black]{hyperref}
\usepackage{authblk}

\providecommand{\keywords}[1]
{
  \small	
  \textbf{\textit{Keywords---}} #1
}

\title{Characterising different communities of Twitter users: Migrants and natives}

\author[1]{Jisu Kim}
\author[2]{Alina S{\^i}rbu}
\author[3]{Fosca Giannotti}
\author[3]{Giulio Rossetti}

\affil[1]{Scuola Normale Superiore, Pisa, Italy.
\url{jisu.kim@sns.it}}
\affil[2]{University of Pisa, Pisa, Itlay.
\url{alina.sirbu@unipi.it}}
\affil[3]{Istituto di Scienza e Tecnologie dell'Informazione, National Research Council of Italy, Pisa, Italy. \url{{fosca.giannotti, giulio.rossetti}@isti.cnr.it}}

\begin{document}

\date{}

\maketitle
\begin{abstract}
    Today, many users are actively using Twitter to express their opinions and to share information. Thanks to the availability of the data, researchers have studied behaviours and social networks of these users. International migration studies have also benefited from this social media platform to improve migration statistics. Although diverse types of social networks have been studied so far on Twitter, social networks of migrants and natives have not been studied before. This paper aims to fill this gap by studying characteristics and behaviours of migrants and natives on Twitter. To do so, we perform a general assessment of features including profiles and tweets, and an extensive network analysis on the network. We find that migrants have more followers than friends. They have also tweeted more despite that both of the groups have similar account ages. More interestingly, the assortativity scores showed that users tend to connect based on nationality more than country of residence, and this is more the case for migrants than natives. Furthermore, both natives and migrants tend to connect mostly with natives.

\end{abstract}

\section{Introduction}
    
    Twitter is one of the microblogging platforms that attracted many users. Unlike some of the other platforms, Twitter is widely used to communicate in  real-time and share news among different users \cite{kwak2010twitter}.
    On Twitter, users follow other accounts that interest them to receive updates on their messages, called ``tweets". 
    Tweets can include photos, GIFS, videos, hashtags and polls. Amongst them, hashtags are widely used to facilitate cross-referencing contents. The tweets can also be retweeted by other users who wish to spread the information among their networks. This involves sometimes adding new information or expressing opinion on the information stated. 
    Despite the limit on maximum 280 characters of tweets\footnote{\url{https://developer.twitter.com/en/docs/counting-characters}}, users are able to effectively communicate with others. 
    
    But above all, Twitter has become a useful resource for research. Twitter data can be accessed freely through an application programming interface (API)\footnote{https://developer.twitter.com/en/docs/twitter-api}. On top of this, the geo-tagged tweets are widely used to analyse real-world behaviours. One of fields of research that makes use of geo-tagged tweets is migration studies. Typically, migration studies have relied on traditional data such as census, survey and register data. However, provided with alternative data sources to study migration statistics in the recent period, many studies have developed new methodologies to complement traditional data sources (See for instance, \cite{kim2020digital,hausmann2018measuring,hawelka2014geo,zagheni2014inferring,mazzoli2020migrant,sirbu2020human}). 
    While these studies have successfully shown advantages of alternative data sources, distinguishing characteristics and behaviours of migrants and natives on Twitter have not been fully understood. 
    
    Here, we aim to study the characteristics and behaviours of two different communities on Twitter: migrants and natives. We plan to do so through a general assessment of features of individual users from profiles and tweets and an extensive network analysis to understand the structure of the different communities.
    For this, we identified 4,940 migrant users and 46,948 native users across 174 countries of origin and 186 countries of residence using the methodology developed by \cite{kim2020digital}.
    For each user, we have their profile information which includes account age, whether the account is a verified account, number of friends, followers and tweets. We also have information extracted from the public tweets which includes language, location (at country level) and hashtags. With these collected data, we explore how each of the communities utilises Twitter and their interests in both the world- and local-level news using the method developed by \cite{kim2021daha}.
    Furthermore, we also explore their social links by studying the properties of the mixed network between migrants and natives. We study centrality and assortativity of the nodes in the network.

    We discovered that migrants tend to have more followers than friends. They also tweet more and from various locations and languages. The assortativity scores show that users tend to connect based on nationality more than country of residence, and this is true more for migrants than natives. Furthermore, both natives and migrants tend to connect mostly with natives.

    The rest of the article is organised as follows: we begin with related works, followed by Section~\ref{sec:data} on data and the identification strategy for labelling migrants and natives on Twitter. Section~\ref{sec:usage} focuses on statistics on different features of Twitter and Section~\ref{sec:network} deals with analysis of the different networks. We then conclude the paper in Section~\ref{sec:concl}. 
    
\section{Related works}

    Many studies exist that analyse different networks on microblogging platforms. Twitter is one of the platforms that has been studied extensively as it enables us to collect directed graphs unlike Facebook for instance. We can study various types of relationships defined by either a friendship (followers or friends\footnote{Followers are users that follow a specific user and friends are  users that a specific user follows. \url{https://developer.twitter.com/en/docs/twitter-api/v1/accounts-and-users/follow-search-get-users/overview}}), conversation threads (tweets and retweets) or semantics (tweets and hashtags). Performing network analysis on these allows us to study properties, structures and dynamics of various types of social relationships.
    
  One of the the first quantitative studies on topological characteristics of Twitter and its role in information sharing is~\cite{kwak2010twitter}. From this study onward, many have found distinguished characteristics of Twitter's social networks. According to the study, Twitter has a ``non-power-law follower distribution, a short effective diameter, and low reciprocity". The study showed that unlike other microblogging platforms that serve as mainly social networking platforms, Twitter acts as a news media platform where users follow others to receive updates on others' tweets. A further study of the power of Twitter in information sharing and role of influencers is~ \cite{cha2010measuring}. The authors focused on three different types of influence: indegree, retweets and mentions of tweets. They found that receiving many in-links does not produce enough evidence for influence of a user but the content of tweets created, including the retweets, mentions and topics matter equally. The same authors extended the work to observe information spreaders on Twitter based on certain properties of the users which led to a natural division into three groups: mass media, grassroots (ordinary users) and evangelists (opinion leaders) \cite{cha2012world}. Furthermore, by looking at the six major topics in 2009 and how these topics circulated, they found different roles played by each group. For example, mass media and evangelists play a major role in spreading new events despite of their small presence. On the other hand, grassroots users act as gossip-like spreaders. The grassroots and evangelists are more involved to form social relationships. 
    
    Studies that appear in the latter years focused on characteristics on Twitter networks and properties in various scenarios, e.g. political context, social movements, urban mobility and more (See for instance \cite{xiong2019hashtag,radicioni2020analysing,mirzaee2020urban}). 
    For instance, \cite{grandjean2016social} studied the network of followers on Twitter in the digital humanities community and showed that linguistic groups are the main drivers to formation of diverse communities. 
    Our work contributes to the same line of these works. But unlike any precedent works, here we explore new types of communities that, to the best of our knowledge, have not yet been explored, i.e., migrants and natives.

\section{Data and labelling strategy}
\label{sec:data}
    \subsection{Data}
    The dataset used in this work is similar to the one used in \cite{kim2021daha}. We begin with Twitter data collected by \cite{coletto2017perception}, from which we extract all geo-tagged tweets from August 2015 to October 2015 published from Italy, resulting in a total of 34,160 individual users (that we call first layer users). We then searched for their friends, i.e. other accounts that first layer users are following which added 258,455 users to the dataset (called second layer users).
    We further augmented our data by scraping also the friends of the 258,455 users. The size of the data grew extensively up to about 60 million users.  To ensure  sufficient number of geo-tagged tweets, all of these users' 200 most recent tweets were also collected. To synthesise the dataset, we focus on a subset of these users for whom we have their social network, and which have published geo-located tweets. This results in total of 200,354 users from the first and second layers with some overlaps present among the two layers.

    \subsection{Labelling migrants and natives}
    
      \begin{figure}[t]
        \centering
        \includegraphics[width=\textwidth]{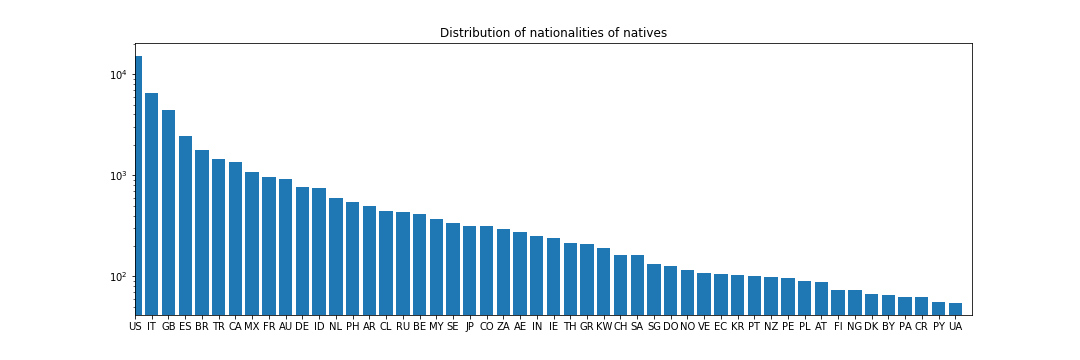}
        \caption{Distribution of top 50 nationalities of natives in log scale}
        \label{fig:dist_nat}
    \end{figure}

       \begin{figure}[t]
        \centering
        \includegraphics[width=0.7\textwidth]{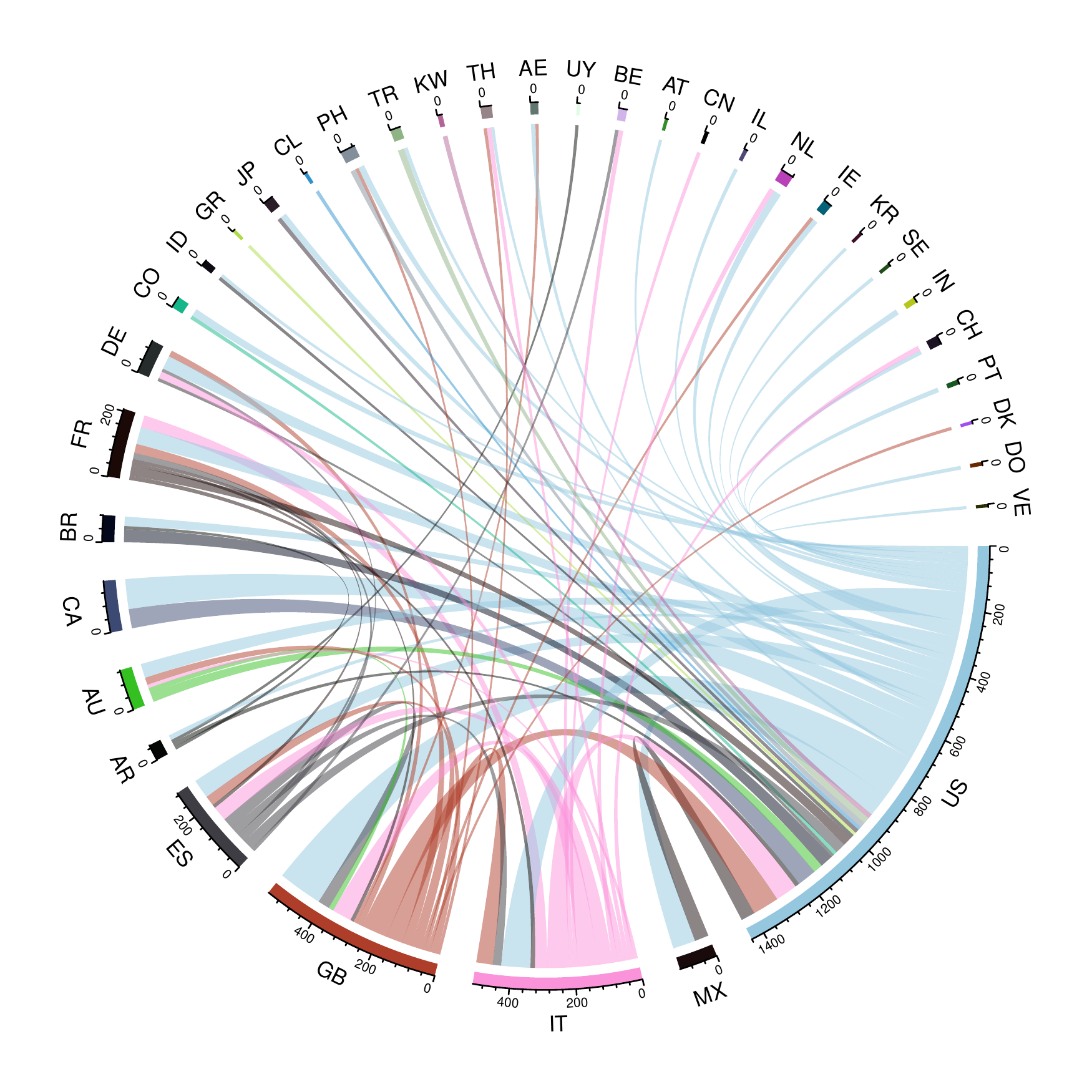}
        \caption{
        Chord diagram on migration patterns: The number of migrants who have moved from a country to another is represented by the links. The colours represent the nationalities of migrants. We show only countries with at least 10 migrants for the visualisation purpose. } 
        \label{fig:chord}
    \end{figure}
    
    The strategy for labelling migrants and natives originates from the work of \cite{kim2020digital}. It involves assigning a country of nationality $C_n(u)$ and a country of residence $C_r(u)$ to each user $u$, for the year 2018. The definition of a migrant is ``a person who has the residence different from the nationality", i.e. $C_n(u) \neq C_r(u)$. 
    The strategy to assign a user's residence requires observing the number of days spent in different countries in 2018 through the time stamps of the tweets. In other words, the country of residence is the location where the user remains most of the time in 2018.
    To assign nationality, we analyse the tweet locations of the user and user's friends. In this work, we took into account the fact that tweet language was not considered important in defining the nationality as found in the study of \cite{kim2020digital}. Thus, the language was not considered here as well.  By comparing the labels of country of residence and the nationality, we determined whether the user was a migrant or a native in 2018.

    Some users could not be labelled since the procedure outlined in \cite{kim2020digital} only assigns labels when enough data is available. As a result, we identified nationalities of 197,464 users and the residence 57,299 users.  Amongst them, the total number of users that have both the nationality and residence labels are 51,888.
    Most importantly, we were able to identify 4,940 migrant users and 46,948 natives from our Twitter dataset. 
    In total, we have identified 163 countries of nationalities for natives.
    From the figure \ref{fig:dist_nat}, we see that the most present countries are the United States of America, Italy, Great Britain and Spain in terms of nationality. 
    This is due to several factors. First because Twitter's main users are from the United States. Second, we have large number of Italian nationalities present due to the fact that we initially selected the users whose geo-tags were from Italy. 
    Figure~\ref{fig:chord} displays the main migration links in our dataset for the countries that have at least 10 migrants. 
    Overall, we have identified 144 countries of nationalities and 169 countries of residences for the migrants. 
    In terms of migration patterns, it is interesting to also remark from our data that the U.S. and U.K have significant number of in and out-going links. In addition, France and Germany have mainly in-coming links.
    
    Here, we emphasise that through our labelling process we do not intend to reflect a global view of the world's migration patterns but simply what is demonstrated through our dataset. However as it is also shown in the work of \cite{kim2021daha}, the predicted data correlate fairly with official data when looking at countries separately. For instance, when comparing predicted data with Italian emigration data of AIRE\footnote{Anagrafe degli italiani residenti all'estero (AIRE) is the Italian register data.}, we observed a correlation coefficient of 0.831 for European countries and 0.56 for non-European countries. When compared with Eurostat data on European countries, the correlation coefficient was 0.762. 
    This provides us the confidence to employ this dataset to analyse characteristics of different communities through Twitter.

\section{Twitter features}
\label{sec:usage}

In this section we look at the way migrants and natives employ Twitter to connect with friends and produce and consume information. 

\subsubsection{Home and Destination Attachment index}
\label{sec:attachment}
A first analysis concentrates on the types of information that users share, from the point of view of the country where the topics are discussed. In particular, we compute two indices  developed by \cite{kim2021daha} : Home Attachment  ($HA$) and Destination Attachment ($DA$), which describe how much users concentrate on topics from they nationality and residence country, respectively. We compute the two indices for both migrants and natives; obviously, for natives the residence and nationality are equal and thus the two indices coincide. 

To compute $HA$ and $DA$, we first assign nationalities to hashtags by considering the most frequent country of residence of natives using the hashtags. A few hashtags are not labelled, if their distribution across countries is heterogeneous (as measured by the entropy of the distribution). 
  The $HA$ is then computed for each user as the proportion of hashtags specific to the country of nationality. Similarly, the $DA$ is the proportion of hashtags specific to the country of residence. Thus, the $HA$ index measures how much a user is interested in what is happening in his/her country of nationality and the $DA$ index reflects how much a user is interested in what is happening in his/her country of residence. 

\begin{figure}
        \centering
        \includegraphics[width=.49\textwidth]{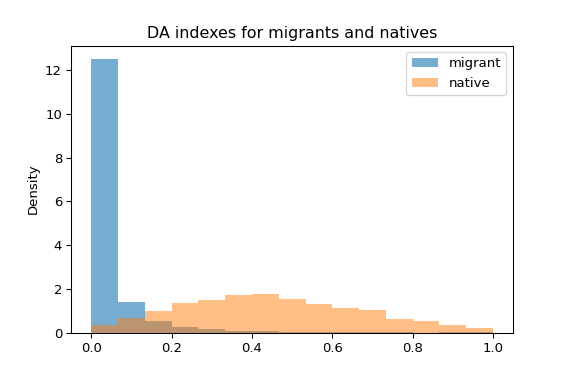}
        \includegraphics[width=.49\textwidth]{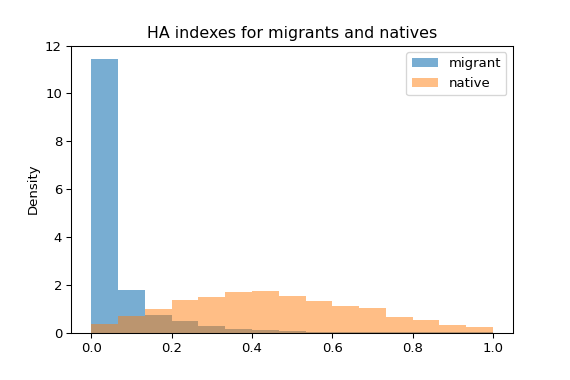}
        \caption{Distribution of DA \& HA for migrants and natives}
        \label{fig:da_ha}
    \end{figure}

    As shown in the figure \ref{fig:da_ha}, the indices clearly behave differently for the two groups: migrants and natives. 
    Similar to \cite{kim2021daha}, we observe that migrants have, on average, very low level of $DA$ and $HA$ . 
    When looking at natives, this index distribution is wider and has an average of 0.447 which is surely higher than the average of migrants. Without a doubt, this shows that natives are more attached to topics of their countries, while migrants are generally less involved in discussing the topics, both for the home and destination country. 
    However, we observe that a few migrant users do have large $HA$ and $DA$ showing different cultural integration patterns, as detailed in~\cite{kim2021daha}. At the same time, some natives show low interest in the country's topics, which could be due to interest in world-level topics rather local-level topics.

\subsection{Profile information}
    
    \begin{figure}[t]
        \centering
        \includegraphics[width=.7\textwidth]{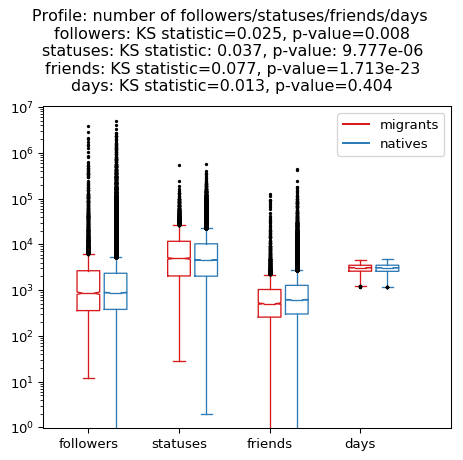}
        \caption{Distributions of profile features: number of days since the account was created untill 2018, number of followers, number of friends, and number of tweets published (statuses).}
        \label{fig:followers_friends}
    \end{figure}
   
    Can we find any distinctive characteristics of migrants and natives from the profiles of users? Here, we look at public information provided by the users themselves on their profiles. We examine the distribution of profile information and perform Kolmogorov–Smirnov (KS) test to compare the distributions for migrants and natives.
    
    On the profile, various information are declared by the users themselves such as the joined date, location, bio, birthday and more. 
    We begin by looking at the age of the Twitter accounts from the moment they created their accounts till 2018, as shown in the figure \ref{fig:followers_friends}. We observe that migrants and natives have similar shape of distributions, providing information that there is no earlier or later arrival of one group or another on Twitter. The KS test with high p-value of 0.404 also confirms that the two distributions are indeed very similar.

    The other criteria we study show some differences. First, we generally observe that natives have slightly more friends than migrants. On average, migrants follow about 1,160 friends and 1,291 friends for the natives. We can also see from the figure \ref{fig:followers_friends} that the range of this number is much wider for the natives, ranging from 0 to maximum of 436,299 whereas for the migrants, this range ends at 125,315. The KS test yields a p-value of 1.713e$^{-23}$, confirming that the two distributions are different. 
    
    Secondly, we observe that the migrants have a larger number of followers. On average, migrants have 10,972 followers versus 7,022 followers for natives (KS p-value of 0.008). This tells us that there are more users on average that are waiting to get updates on migrant users' tweets. 
    Interestingly, when it comes to the number of tweets (statuses) that users have ever tweeted since the account was created, the number is about 9\% higher for the migrants than the natives: average values of 9,836 for migrants and 9,016 for natives, p-value of 9.777e$^{-06}$. 
    
    We also look at the number of accounts that are classified as verified accounts. The verified accounts are usually well-known people such as celebrities, politicians, writers, or directors and so on. Indeed when looking at the proportion of verified accounts, we observe that this proportion is higher among migrants than natives which partly explains also the higher number of followers and tweets for this group. To be more specific, 5\% of the users' accounts are verified accounts among migrants and 3.7\% of the accounts are verified accounts among natives.

\subsection{Tweets}
    
\begin{figure}
        \centering
        \includegraphics[width=.7\textwidth]{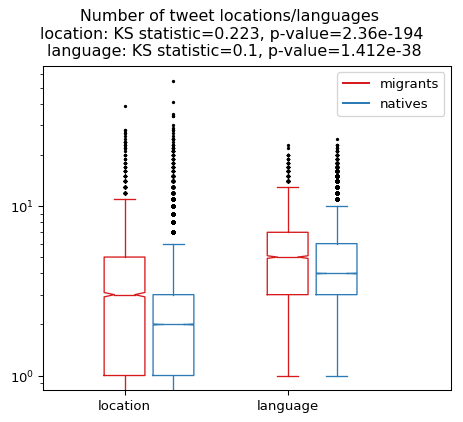}
        \caption{Distribution of tweet locations and languages}
        \label{fig:tw_loc_lang}
    \end{figure}

    Tweets also provide useful information about user behaviour. We are interested in the locations (country level) and languages a user employs on Twitter. Hence, we look at the number of languages and locations that appear in the users' 200 most recent tweets and computed also the KS statistics to compare the differences between the distributions of migrants and natives. 
    As shown in Figure~\ref{fig:tw_loc_lang}, we note that migrants tweet in a wider variety of languages and locations. The two distributions for migrants and natives are different from each other as the KS tests show low p-values; 2.36e$^{-194}$ for location and 1.412e$^{-38}$ for language.
    
\begin{figure}
        \centering
        \includegraphics[width=.7\textwidth]{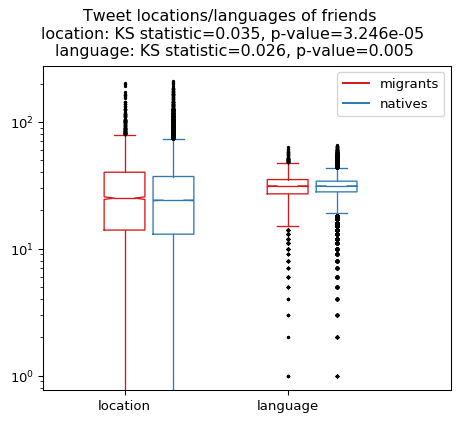}
        \caption{Distribution of tweet locations and languages of friends}
        \label{fig:tw_loc_lang_fr}
    \end{figure}

    Since we possess network information, we also studied the tweet language and location information for a user's friends. In Figure~\ref{fig:tw_loc_lang_fr}, the two distributions show smaller differences among natives and migrants, compared to Figure~\ref{fig:tw_loc_lang}. However, the p-value of the KS test tells us that the distributions are indeed different from one another, where the p-value for location and language distribution for migrants and natives are 3.246e$^{-05}$ and 0.005 respectively. Although the differences are small , we observe that the friends of migrants tweet in more numerous locations than those of natives, with average of 29.6 for migrants and 27.4 for natives.
    However, although the two distributions are different from each other from the KS p-value, the actual difference between average values is very small in the case of the number of languages of friends. In fact, the average for migrants is 30.22 and 30.43 for natives.

    These numbers indicate that the migrants have travelled in more various places and hence write in diverse languages than the natives. The friends of migrants tend to have travelled more also. However, no large differences were observed for the number of languages that friends can write in for both migrants and natives.

\subsubsection{Popular hashtags}
\begin{figure}
        \centering
        \includegraphics[width=0.7\textwidth]{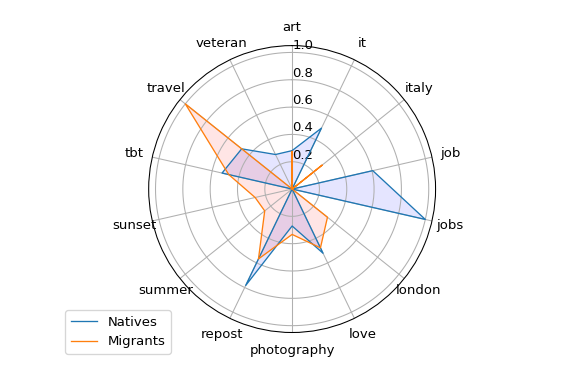}
        \caption{Top 10 hashtags used by migrants and natives}
        \label{fig:web}
    \end{figure}

    What were the most popular hashtags used by natives and migrants in 2018? In Figure~\ref{fig:web} we display the top 10 hashtags used by the two communities, together with the number of tweets using those hashtags, scaled to $[0,1]$. We observe  that natives and migrants share some common interests but they also have differences. For instance, some of the common hashtags between natives and migrants are \#tbt, \#love and \#art. Other hashtags such as \#travel, and \#repost are in the top list but the usage of these hashtags is much higher in one of the groups than the other. For instance, the hashtag \#travel is much more used by migrants than the natives. This is interesting because the number of tweet locations of migrants also reflect their tendency to travel, more than natives. Followed by the hashtag \#travel, migrants also used other hashtags such as \#sunset, \#photography, \#summer, and hashtags for countries which show their interests in travelling. On the other hand, natives are more focused on hashtags such as \#job, \#jobs, and \#veteran.

\section{Network analysis}
\label{sec:network}    

In this section we perform social network analysis on the social graph of our users to examine the relationships between and within the different communities, i.e., migrants, and natives. Initially, our network consisted of 45,348 nodes and 232,000 edges. We however focus on the giant component of the network which consists of 44,582 nodes and 231,372 edges. Each node represents either a migrant or a native and the edges are directed and represent friendship on Twitter (in other words, our source nodes are following the target nodes). Since we have migrants and natives labels, our network allows us to study the relationship between migrants and natives. 

\subsection{Properties of the network}

        \begin{figure}[]
        \centering
        \includegraphics[width=.7\textwidth]{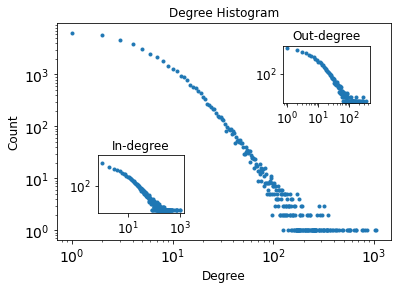}
        \caption{Degree distribution of the network.}
        \label{fig:degree}
    \end{figure}
    
    In this section we start by looking at density, reciprocity, and shortest path length for the network, and then study node centrality including degree distribution. The average density score of our network tells us that on average each node is connected to other 5.2 nodes.  
    The reciprocity coefficient is low and indicates that only 23.8\% of our nodes are mutually linked. This is normal on Twitter as most of the users follow celebrities but the other way around does not happen in many cases. 
    Within the network, the average shortest path length is 2.42, which means we need on average almost 3 hops to receive information from one node to another.

    \begin{figure}
        \centering
        \includegraphics[width=.49\textwidth]{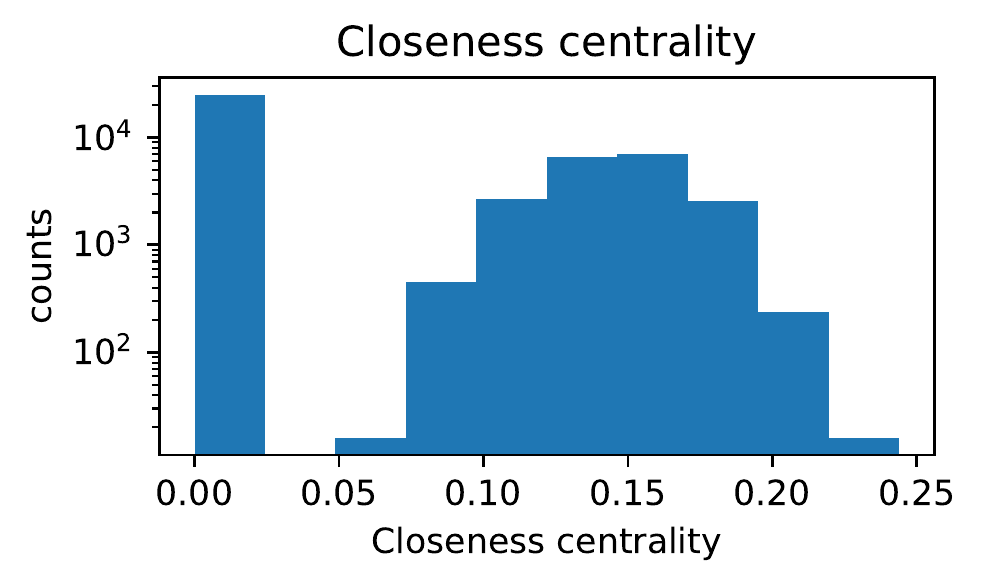}
        \includegraphics[width=.49\textwidth]{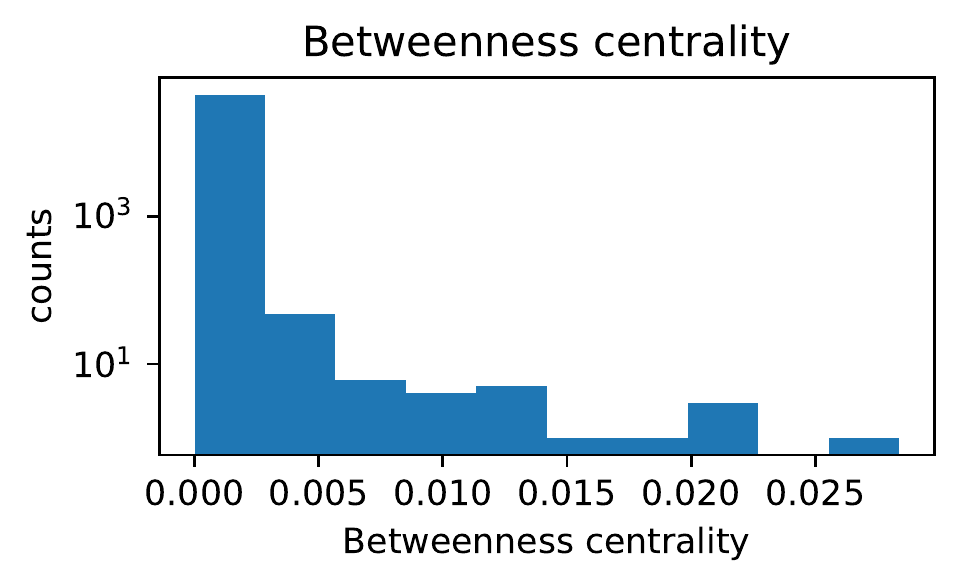}
        \includegraphics[width=.49\textwidth]{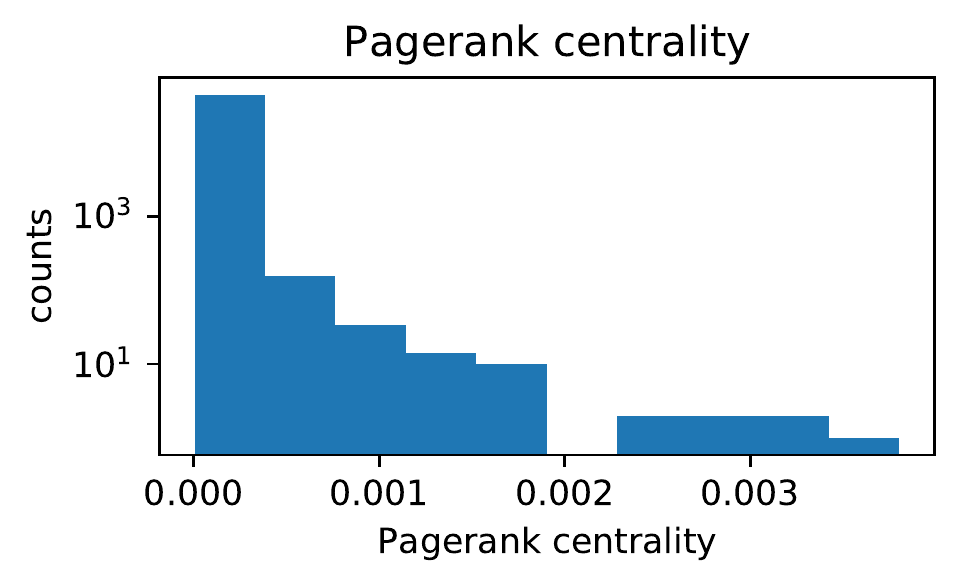}
        \includegraphics[width=.49
        \textwidth]{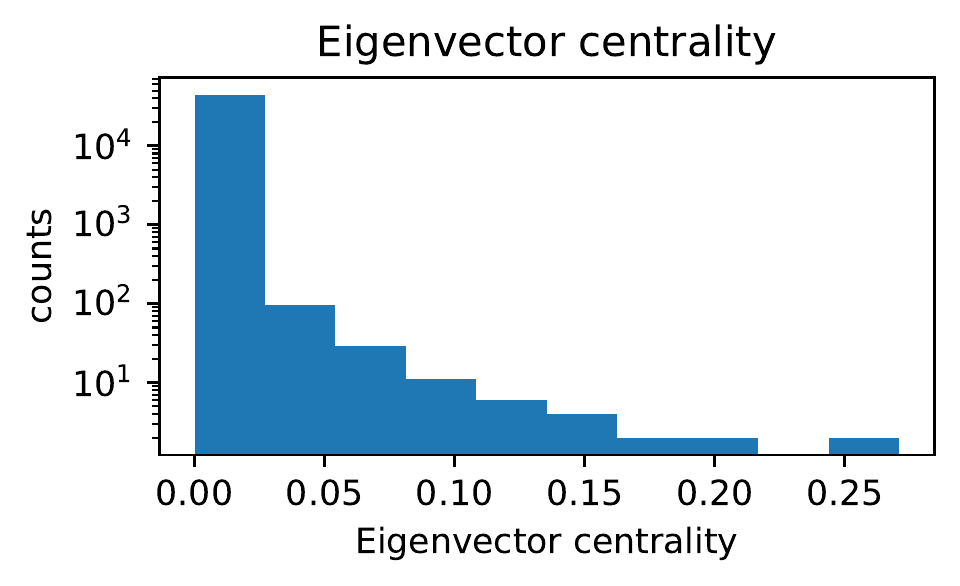}
        \caption{Centrality measures of the network.}
        \label{fig:centrality}
    \end{figure}

    We also compute 7 measures of centrality to study. The measures include all-, in- and out-Degree (Figure~\ref{fig:degree}) plus Closeness,  Betweenness, Pagerank and Eigenvector centrality measures (shown in Figure~\ref{fig:centrality}).
    The Degree centrality measures the number of connections that a particular node has, which can either be an in- or out- going connection. The Pagerank measure considers that nodes with low out-degree are more important.
    The Betweenness centrality looks at nodes that serve as a bridge from one part of a graph to another. 
    On the other hand, the Closeness centrality looks at how the node is in a most favourable position to control and acquire vital information within the network.
    Lastly, the Eigenvector measure considers that a node is important if the node is connected to other highly connected nodes.

\begin{figure}
    \centering
    \includegraphics[width=\textwidth]{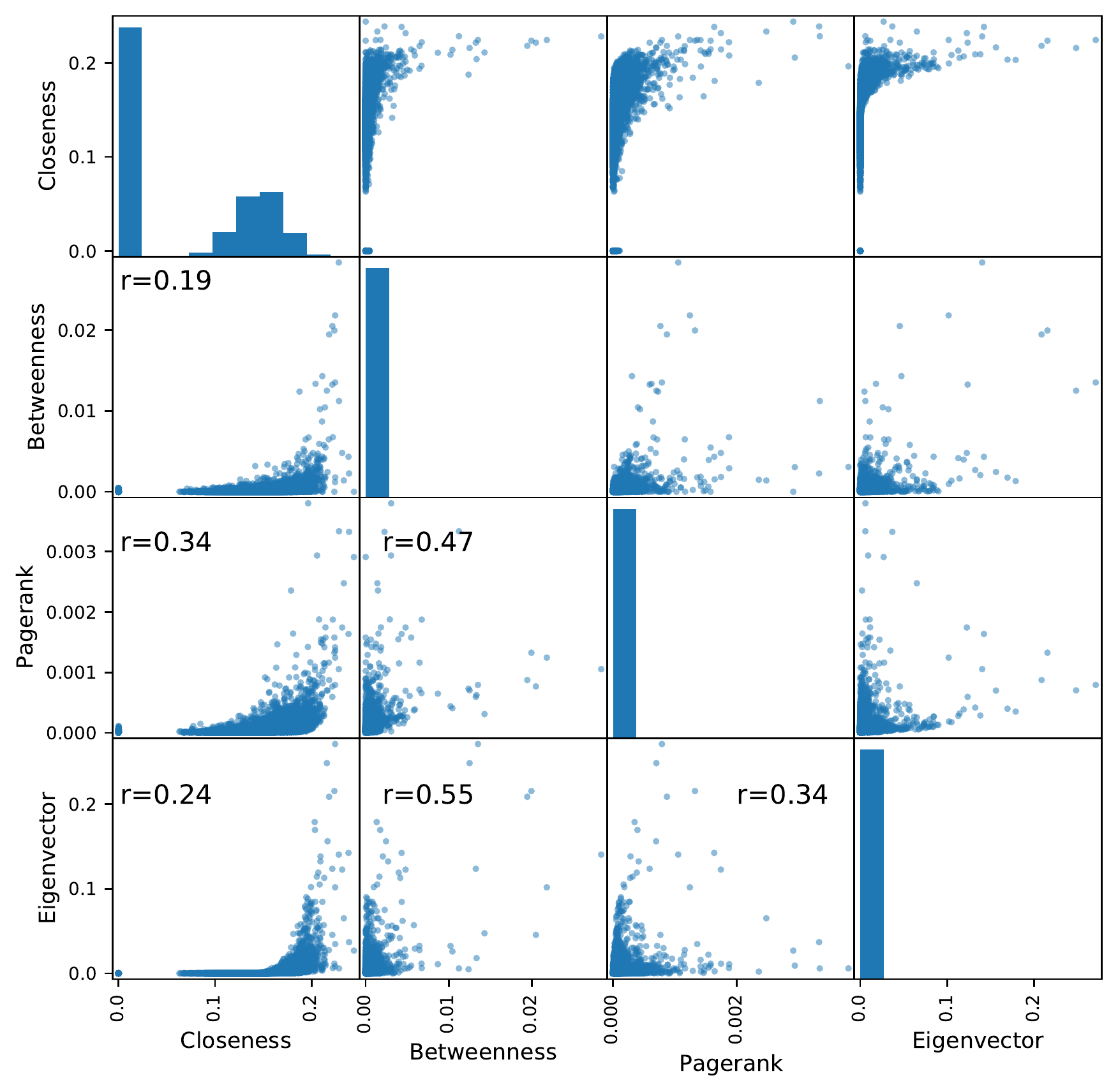}
    \caption{Correlation between different centrality measures for network}
    \label{fig:corr_network}
\end{figure}    

\begin{figure}
    \centering
    \includegraphics[width=0.8\textwidth]{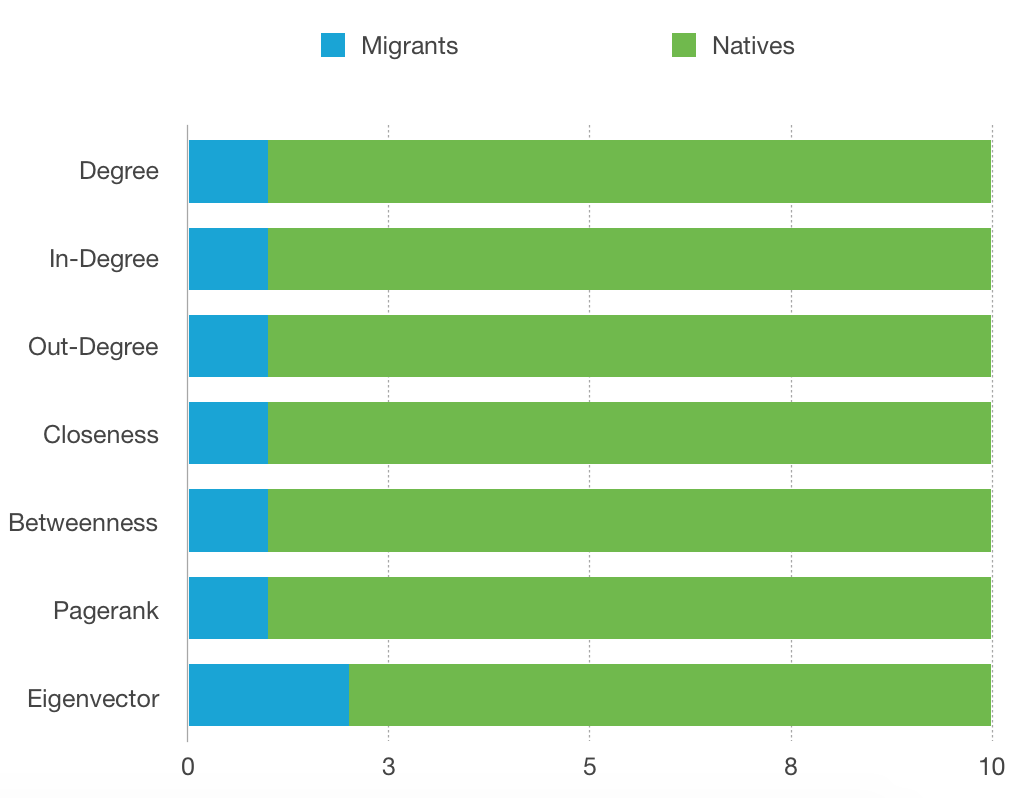}
    \caption{Summary of labels of users for top 10 central users by different centrality measures.}
    \label{fig:central}
\end{figure}

     As we can observe from Figure~\ref{fig:degree}, the degree distribution follows a power-law distribution with alpha equal to 2.9. This means that a minority of the nodes are highly connected to the rest of the nodes.
    From Figure~\ref{fig:centrality}, we observe that most of the users have low centrality while a small number of users show higher centrality values. This is true for all measures, however for closeness the number of users who show higher centrality is larger than for the other measures.
   This means that many users are well-embedded in the core of the network, and are in a good position to receive information. 
    The distribution of Betweenness, on the other hand, tells us that small part of the users are situated in the most crucial points in case of a diffusion process. We also note that that range of betweenness values is rather narrow, even users with the largest betweenness show a small  value. This indicates that information is flowing rather uniformly through the network, and no nodes are particularly important in the process. This was also shown previously by the low average shortest path length.  
    A similar situation arises with the Pagerank measure of centrality: a small minority of users show higher values, however they are all rather low which indicates that generally the network is quite uniform.
    Lastly, the Eigenvector centrality reveals that a small part of users has influence even beyond the nodes that are directly connected to. Overall, the centrality measures seem to indicate that while the topological structure of the network is heterogeneous with some nodes showing higher connection and centrality, from the point of view of the flow of information the user in our networks have similar roles. 
    
     We continue to examine the centrality of users by computing the correlation between the different measures as shown in Figure~\ref{fig:corr_network}. First of all, we observe a positive relationship among all measures, which is expected, as it means that users who are central from one point of view are also central from another. The Betweenness and Eigenvector centrality measures correlate the most (r=0.55). This tells us that users that serve as a bridge between two parts of graphs are also likely to be the most influential user in the network.
    On the other hand, Betweenness and Closeness centrality measures have the lowest correlation with r=0.19. However, the scatterplot shows that those few users who have larger Betweenness also have a large Closeness. The low correlation is determined by the fact that a large majority of users show almost null Betweenness, however Closeness is heterogeneous among this group. A similar observation can be made for the relation between Closeness on one side and Pagerank and Eigenvector centrality on the other: high Pagerank and Eigenvector centralities always correspond to high Closeness, however for users with low Pagerank and Eigenvector centrality the Closeness values vary. 
    
    When checking the labels, in terms of migrant or native, of the most central users, we see that in general these are mostly natives.
    To be more specific, in Figure~\ref{fig:central} we show the labels of the top 10 users for each centrality measure. We observe that among the top 10, 8 or 9 users are natives. In other words, most of the nodes have majority of in- and out-going links directed to natives' accounts. This is somewhat expected since in our network only 10\% of users are migrants. However, we note that a migrant user is always in the top 3 in Closeness, Pagerank and Eigenvector centrality measures. This tells us that this migrant user has a crucial influence over the network around itself but also beyond its connections.

\subsection{Assortativity analysis}

    We now focus on measuring assortativity of nodes by different attributes of individuals, i.e., migrants or natives, country of residence and country of nationality. Assortativity tells us whether the network connections correlate in any way with the given node attributes. In our case this analysis allows us to infer whether and in what measure the network topology follows the nationality or residence of the users, or whether the migrant/native status is relevant when building online social links.
    
    We begin with global assortativity measures. First, the degree assortativity coefficient of -0.046 shows no particular homophily behaviour from the point of view of the node degree. That means high degree nodes do not link with other high degree nodes.
    However, when we measure the assortativity by different attributes, we observe different results.
    When looking at the coefficient by the country of residence, the score of 0.55 shows a very good homophily level. The score improves slightly when we examine the behaviour through the attributes of country of nationality (0.6).
    These values tell us that nodes tend to follow other nodes that share same country of residence and country of nationality, with a stronger effect for the latter.
    However, when looking at the coefficient by the migrant/native label, we observe no particular correlation (0.037).
    
\begin{figure}
        \centering
        \includegraphics[width=.49\textwidth]{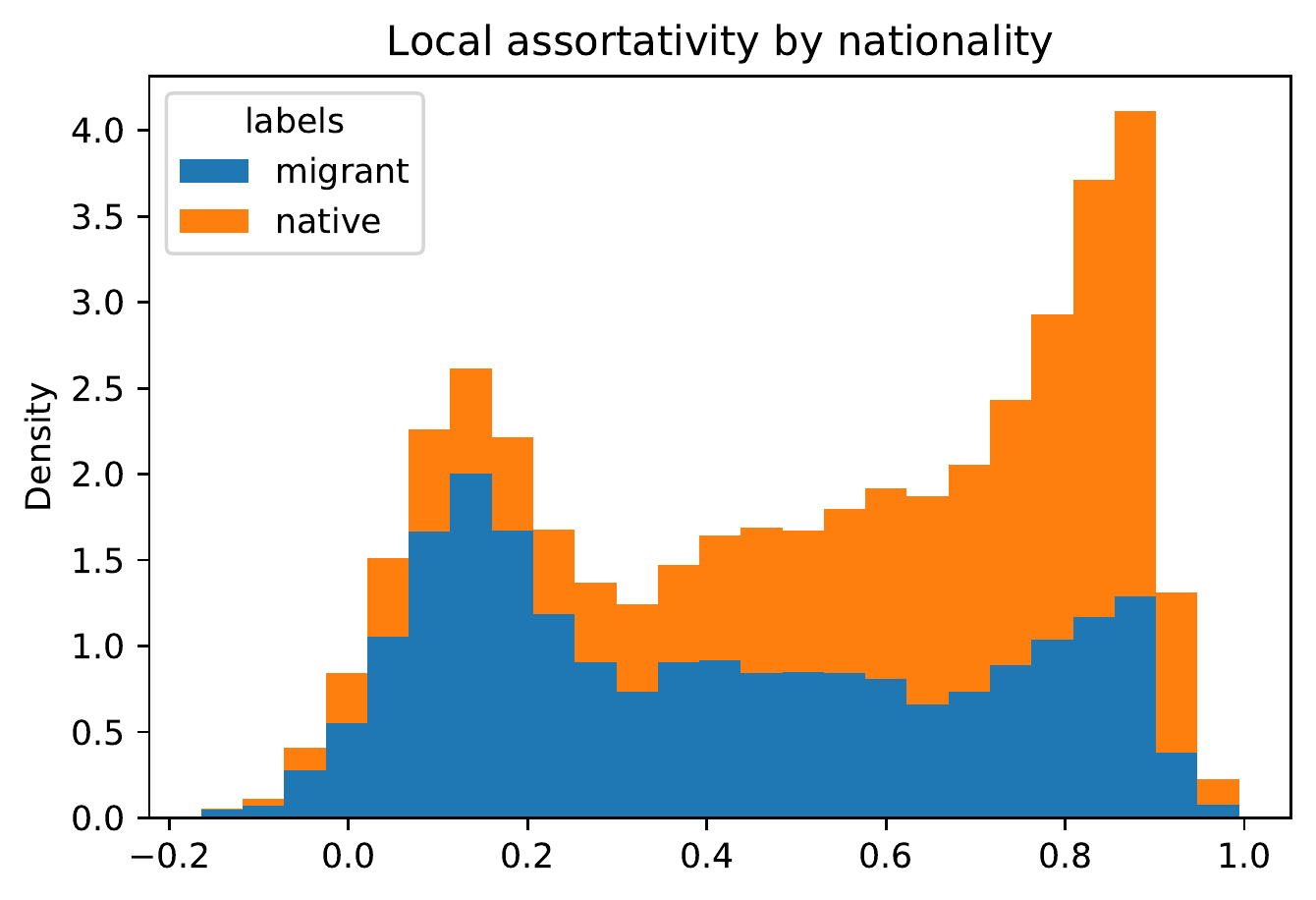}
        \includegraphics[width=.49\textwidth]{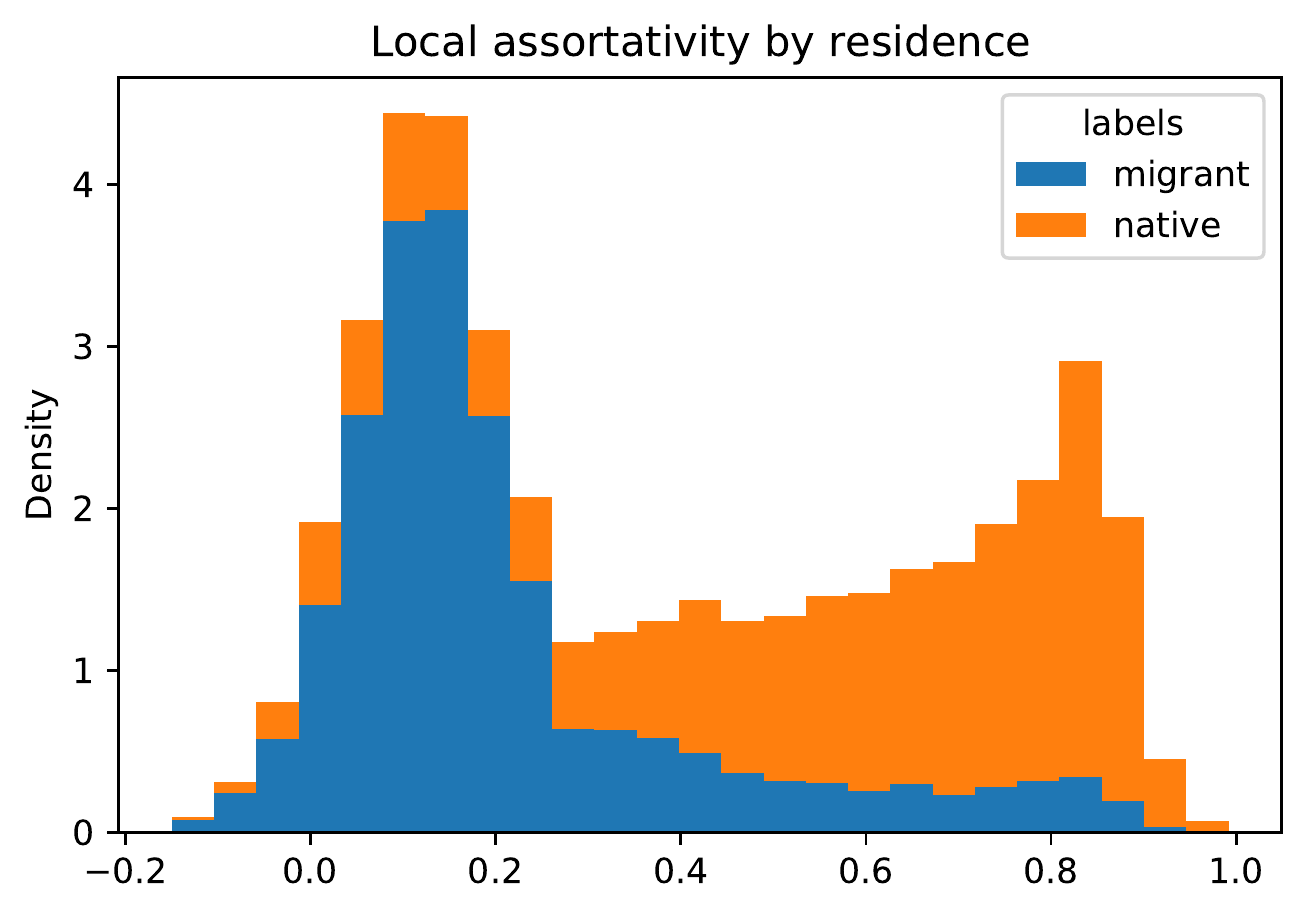}
        \includegraphics[width=.49\textwidth]{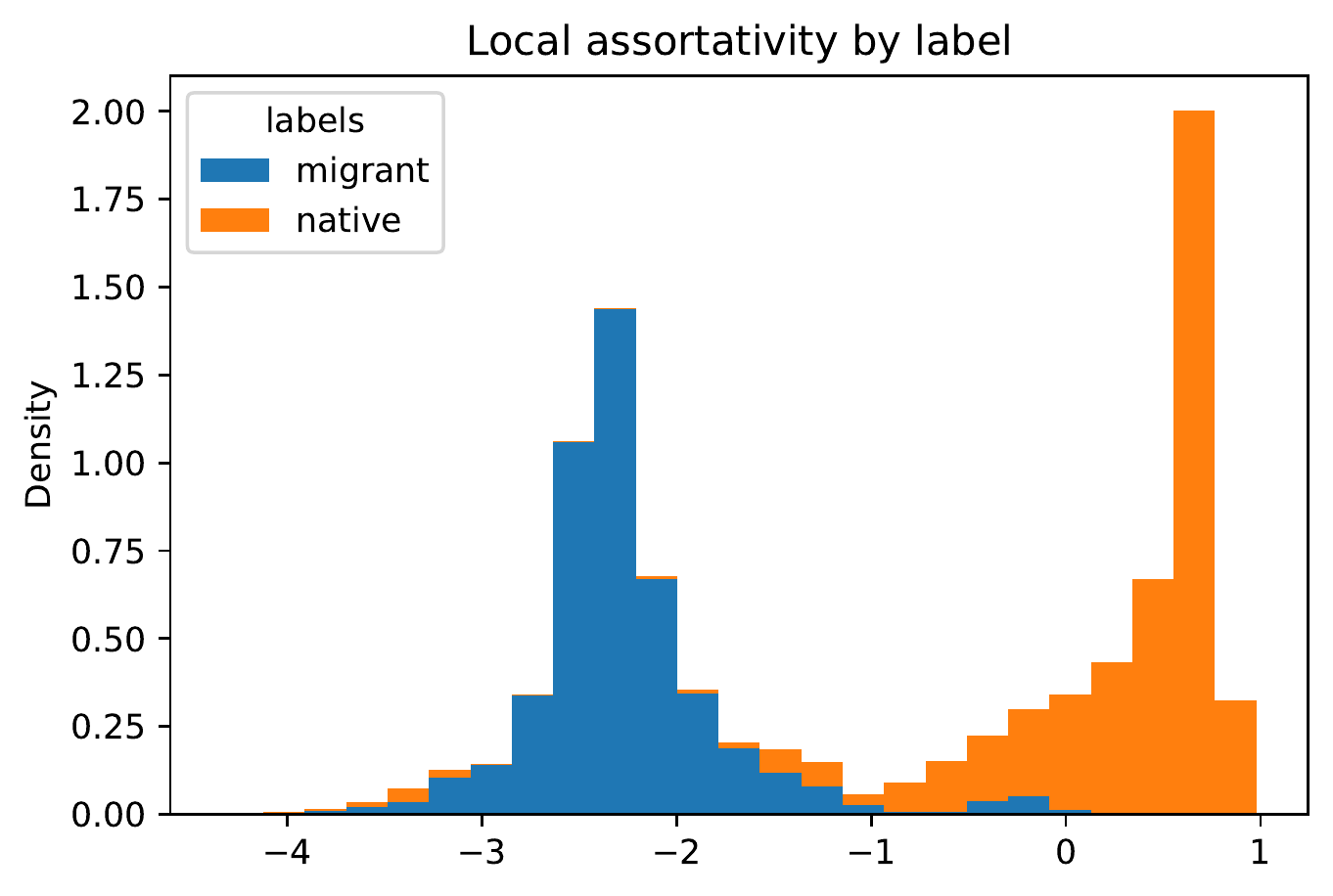}
        \caption{Stacked histogram of local assortativity: From the top we have local assortativity by nationality, by residence and by migrant/native label. Please note that the histograms are stacked, therefore there is no overlap between the plot bars.}
        \label{fig:assort}
    \end{figure}

    The global assortativity scores are susceptible to be influenced by the size of the data and the imbalance in labels, which is our case especially for the migrant/native labels. Therefore we continue to examine the assortativity at local level, allowing us to overcome the possible issues at global level. We thus compute the scores based on an extension of Newman's assortativity introduced by~\cite{peel2018multiscale}. In Figure~\ref{fig:assort} we show the distribution of node-level assortativity of migrants and natives, for the three attributes (nationality, residence and migrant/native label). We observe again good homophily for all attributes at local level. 
    However, we remark different behaviour patterns for migrants and natives.
    Specifically, we see that migrants tend to display lower homophily compared to natives, when looking at the assortativity of nodes by country of residence and migrant/native labels. 
    This tells us that migrant users tend to consider less the country of residence when following other users. Instead, most natives tent to connect with users residing in the same country.
    When looking at nationality, this effect is less pronounced. While natives continue to display generally high homophily, with a small proportion of users with low values, migrants show a flatter distribution compared to the nationality. Again, a large part of migrants show low homophily, however a consistent fraction of migrant users show higher nationality homophily, as oposed to what we saw for the residence. This confirms what we observed at global level: there is a stronger tendency to follow nationality labels when creating social links. 
    As for the assortativity of nodes by migrant/native labels, we observe that migrants and natives clearly have distinctive behaviours. While natives tend to form connections with other natives, migrants tend to connect with natives as well. This could also be due to the fact that migrants are only about 10\% of our users so naturally many friends will be natives (from either residence, nationality or other country).
    This result is different from what we observed at global level and confirms that the global assortativity score was influenced by the size of the data and the imbalance in labels.
    
\section{Conclusion}
\label{sec:concl}
    We studied the characteristics of two different communities; migrants and natives observed on Twitter. Analysing profiles, tweets and network structure of these communities allowed us to discover interesting differences. 
    We observed that migrants have more followers than friends. They also tweet more often and in more various locations and languages. This is also shown through the hashtags where the most popular hashtags used among migrants reflect their interests in travels. 
    Furthermore, we detected that Twitter users tend to be connected to other users that share the same nationality more than the country of residence. This tendency was relatively stronger for migrants than for natives. Furthermore, both natives and migrants tend to connect mostly with natives.

    As mentioned previously, we do not intend to generalise the findings of this work as only a small sample of individual Twitter data was used. 
    However, we believe that by aggregating the individual level data, we were able to extract information that is worthwhile to be investigated further. 
    To this extent, we simply intend to present what is demonstrated through out dataset. 
    In spite of this drawback, we were able to observe social interactions between migrants and natives thanks to the availability of the Twitter data. In the future, we plan to analyse semantic networks of these users' tweets and hashtags to understand core interests of their discussions and how each community gets involved in different discussions.

    \section*{Acknowledgements}
    This work was supported by the European Commission through the Horizon2020 European projects ``SoBigData++: European Integrated Infrastructure for Social Mining and Big Data Analytics" (grant agreement no 871042) and ``HumMingBird  - Enhanced migration measures from a multidimensional perspective" (grant agreement no 870661).
 

\bibliographystyle{plain}
\bibliography{reference}
 \end{document}